\documentclass[11pt]{article}
\pdfoutput=1
\usepackage{amsfonts, amssymb, amsmath, graphicx, epsfig, lscape}
\usepackage{subfigure}
\usepackage{ifpdf}
\newcommand{\qed}{$\Box$}

  {\hspace*{\fill}$\Box$\par\vspace{4mm}}
  {\hspace*{\fill}$\Box$\par\vspace{4mm}}
  {\hspace*{\fill}\par\vspace{4mm}}

\begin {document}

\title{Choreography in Inter-Organizational Innovation Networks}

\author{%
Giovanna Ferraro \and Antonio Iovanella}

\date{}

\maketitle

\begin{center}
{\footnotesize Dipartimento di Ingegneria dell'Impresa\\
University of Rome ``Tor Vergata''\\
Via del Politecnico, 1 - 00133 Rome, Italy.\\
\texttt{[giovanna.ferraro, antonio.iovanella]@uniroma2.it}}
\end{center}

\begin{abstract}
This paper introduces the concept of choreography with respect to inter-organizational innovation networks, as they constitute an attractive environment to create innovation in different sectors. We argue that choreography governs behaviours by shaping the level of connectivity and cohesion among network members. It represents a valid organizational system able to sustain some activities and to reach effects generating innovation outcomes.
This issue is tackled introducing a new framework in which we propose a network model as prerequisite for our hypothesis. The analysis is focused on inter-organizational innovation networks characterized by the presence of hubs, semi-peripheral and peripheral members lacking hierarchical authority.
We sustain that the features of a network, bringing to synchronization phenomena, are extremely similar to those existing in innovation network characterized by the emergence of choreography.
The effectiveness of our model is verified by providing a real case study that gives preliminary empirical hints on the network aptitude to perform choreography. Indeed, the innovation network analysed in the case study reveals characteristics causing synchronization and consequently the establishment of choreography. 

\vspace{5 mm}

\end{abstract}

%%%%%%%%%%%%%%%%%% Introduction %%%%%%%%%%%%%%%%%%

\section{Introduction}
\label{Introduction}

The purpose of this paper is to describe a new framework in which we extend the role of network coordination, direction, influence and management among all members, thus proposing the emergence of choreography. Network structures are well described by the theory of complex networks~\cite{WC2} consisting of nodes or vertices connected by links or edges. They are currently studied across many fields of sciences, technologies and social disciplines~\cite{Est}.

The world choreography is derived from the Greek for ``dance'' $\chi o \rho \epsilon \iota \alpha$ and ``write'' 
$\gamma \rho \alpha \phi i \alpha$ reflecting the sequence of steps and movements in dance, the art or the practice of designing choreographic sequences. The emergence of choreography in a network leads to the establishment of coordinated activities among all members that allows creation and extraction of innovation through the final outcomes~\cite{FI1}.

We examine the inter-organizational innovation networks characterized by recurring exchange interactions among members that retain residual control of their individual resources yet periodically jointly decide over their use. Members of such networks can be firms, organizations or research centres, located in different regions and specialized in particular sectors, linked by common interests, technologies and skills and networked by the decision to collaborate according to specific rules. Technological districts, business incubators and consortia created by international initiatives financed by the European Commission are some examples of such kind of networks.

Innovation networks are defined as: {\em ``a basic institutional arrangement to cope with systemic innovation. Networks can be viewed as an inter-penetrated form of market and organization} [\ldots].  {\em They include joint ventures, licensing arrangements, management contracts, sub-contracting, production sharing and R\&D collaboration''} (\cite{Fre}, \cite{IB}). 

Inter-organizational innovation networks exhibit the presence of members that act as hubs, i.e. nodes to which most of the others link to, as semi-peripheral members that make a relevant contributions in getting great portion of the network together and as peripheral members that are connectors for local portion of the system.

The establishment of choreography requires some structural hypothesis on network topology as well as some network characteristics and involves the accomplishment of certain activities among network members, namely management of knowledge flow, innovation appropriability, management of stability and management of vitality and health~\cite{FI1}. Such activities are similar to the tasks foreseen by the orchestration theories (\cite{DP}, \cite{Hag}, \cite{NS1}, \cite{NS2}, \cite{RAB}, and references therein) with the difference that they consider a single network hub accomplishing such tasks, while in our model they are the consequence of both the network structure and the characteristics of the membership.

In the orchestration literature, the hub executes, by the means of its degree and centrality~\cite{Row}, a set of deliberate and purposeful actions enabling network organization and coordination to create and extract value from the network~\cite{DP} and to reach the innovation outcome.

In this paper, we focus on network structure and network membership as the framework necessary to support choreography. Activities, innovation leverage and coherence and innovation outcome related to choreography are presented in~\cite{FI1}.

The paper is organized as follows: Section \ref{nc} introduces the concept of network choreography; Section \ref{fnc} describes the proposed framework for the establishment of network choreography; Section \ref{EEN} shows the case study that gives preliminary indications on the network attitude to establish choreography; Section \ref{cfd} presents the final thoughts and sums up some of the key issues that require further research.

\section{Network choreography}
\label{nc}

The definition of network choreography is introduced taking into consideration some suggestions coming from the network orchestration, the Agent Theory [24] and the Business Process Modelling (BPM)~\cite{DW}.

In network theory, as well as in Agent Theory or in BPM, the term orchestration refers to the capacity of a particular agent to influence the evolution of a network (\cite{DW}, \cite{RAB}, \cite{ZK}). Thus, orchestration network relates to intra-organization coordination of activities within roles [8] and is focused on internal behaviour of processes~\cite{PZ}.

We focus on network structure introducing the term network choreography as {\em the network's capacity to address collaboration among multiple members}.

Interactions in choreography are concrete instances of activities carried out by members that comply with network rules and purposes. Activities can be among members, for example in case of knowledge flow when two or more agents share information, or can be within agents, where such internal activities are not visible in the network.

In orchestrated networks, activities are implemented according to the members' role in the network and constrained through the corresponding patterns of network composition. Such activities are undertaken through instances of orchestration processes achieved by the hub~\cite{NS2}. The orchestrator, i.e. the hub acts as controller and executer and performs as leader in the network with the possibility to invoke processes execution from all members.

Choreography, on the other hand, is focused on inter-organization coordination for external perspectives~\cite{PZ} while collaboration is addressed by self-organizing interactions. Choreography entails a set of complex interactions among roles that are in executed a peer-to-peer approach; processes are implemented according to network organization.

\begin{table}[htbp]
\begin{small}
\begin{center}
\begin{tabular}{ | l | l |}
\hline
\textbf{Orchestration} & \textbf{Choreography}\\
\hline\hline
Specification of the role patterns&Specification of the collaboration patterns\\
Process model&Interaction model\\
Processes explicitly invoked  &Processes information driven\\
Within single participant&Among participants\\
Centralized controlling&Distributed controlling\\
Centralized executer&Peer-to-peer approach\\
Node focused&Network focused\\
    \hline
    \end{tabular}
    \caption{Orchestration vs. Choreography in networks.}\label{choreography}
\end{center}
\end{small}
\end{table}

Table~\ref{choreography} resumes the main differences between orchestration and choreography. Note that the listed properties for orchestrated networks hold also for orchestrated agents~\cite{ZK}. Therefore, the same similarity can be proposed between network choreography and agents' choreography.

%%%%%%%%%%%%%%%%%% A framework supporting network choreography %%%%%%%%%%%%%%%%%%

\section{A framework supporting network choreography}
\label{fnc}

We introduce a first attempt to describe a framework for innovation networks that permits the emergence of choreography. 
The establishment of choreography requires some structural hypothesis on network topology and some membership characteristics and involves the accomplishment of certain activities among members to reach the innovation outcome. 

\subsection{Network properties}
\label{np}

We propose complex networks with the scale-free feature (\cite{WC2} and references therein) an adequate framework within which to represent real inter-organizational innovation networks.  

\subsubsection{Network topology}
\label{nt}

The mathematical abstraction of complex networks is a graph $G$. A graph $G(V, E)$ is composed of a set of $N$ nodes $V$, a set of $M$ arcs $E$ defining a relationship among these nodes. We refer to a member by an index $i$ meaning that we allow a one-to-one correspondence between an index and a member. We associate to each member a node of a graph: two members, say $i$ and $j$, are adjacent through an arc if and only if there is a connection between $i$ to $j$.  In the following, we call arcs and links indifferently.

A graph can be represented by an adjacency matrix $A$, that is a $N$-square matrix where an element $a_{ij}$ is zero if there are no arcs between nodes $i$ and $j$, otherwise is equal to the number of arcs. The number of arcs connected to a node $i$ is called its degree $k_i$ and can be interpreted in terms of the size of members' neighbourhoods within the network. Since the said graph represents reciprocal relationships, the matrix is symmetric with zero elements on the main diagonal.

Complex networks overcome the limits of random graphs and their unsuitability to denote real systems. Random graphs are represented by the Erd\H{o}s and R\'enyi model~\cite{ER} in which the probability $P$ to have a link between two nodes is a constant divided by $N$. The degree distribution, i.e. the spread of node degree over the network, follows a Poisson distribution, so most nodes have a degree close to the average and the slope of the curve falls more than exponentially. This means that extremely few nodes are very highly connected~\cite{Bar}.

In this paper, we sustain that a suitable model to represent inter-organi-zational innovation networks topology, among the various complex network organizations, is the Barab\'asi-Albert (BA) scale-free network model~\cite{BA}.

Scale-free networks are open and dynamically formed by continuous addition of new nodes that represent members, while links among members mimic collaborative agreements. Links' inhomogeneity reflects the degree of members' involvement in the network and the difference among hubs, semi-peripheral and peripheral members. Scale-free networks emerge in real situations in which the kind of inhomogeneity in the degree distribution represents few nodes having many links whereas the majority of them has few connections. Therefore, a single node or hub cannot be considered representative since these networks are held together by a different, although limited, number of highly connected hubs while the majority of nodes has smaller connection degree than the average. All nodes are linked with a rather short path due to the {\em small world} characteristic~\cite{WC2} even in case of large and sparse systems.

Such networks are mainly identifiable by three characteristics: the average path length, the clustering coefficient and the degree distribution.

The distance $d_{ij}$ between two generic nodes labelled $i$ and $j$, represents the number of links along the shortest path connecting them; the average path length $L$ of a network is the distance between two nodes, averaged over all pairs of nodes determining the size of a network. In general, the average path length of real complex networks is relatively small reflecting the {\em small world} property since it is related to the possibility to link any two nodes in the network through few links.

The clustering coefficient $C$ is the average fraction of neighbours' pairs of nodes (and they are therefore also neighbours of each other) and shows dense local subgroups of interconnected nodes. The clustering coefficient $C_i$ of a node $i$ is defined as:

$$
C_i = \frac{2E_i}{k_i(k_i - 1)}
$$
\noindent
where $E_i$ is the number of links between the neighbours of $i$. The coefficient $C$ of the network is the average of all $C_i$ over all $i$. Note that $C \leq 1$ and it assumes value equal to $1$ in case of a clique, i.e. a fully coupled network.

Clustering is introduced by structural embeddedness that is the existence of dense ties among nodes. In innovation networks, members sharing common partners have knowledge about each other's trustworthiness, capabilities, competences and reputation and thus mitigating the effects of power asymmetries~\cite{CJ}.

Regarding the degree distribution, the degree $k_i$ of a node $i$ represents the number of its links; the larger the degree, the more significant the node is in a network. The average of $k_i$ over all $i$ is the network average degree and the spread of node degrees over a network is characterized by a distribution function $P(k)$ that represents the probability that a randomly selected node has exactly $k$ links.

In BA model, the connectivity distribution follows a power law function, i.e. the probability $P(k)$ that a node in the network interacts with $k$ other vertices decays with the law $P(k) \sim k^{-\gamma}$ with slope $\gamma$ as the scaling exponent. 

\begin{figure}
\centering
\subfigure
	{\includegraphics[scale=.6]{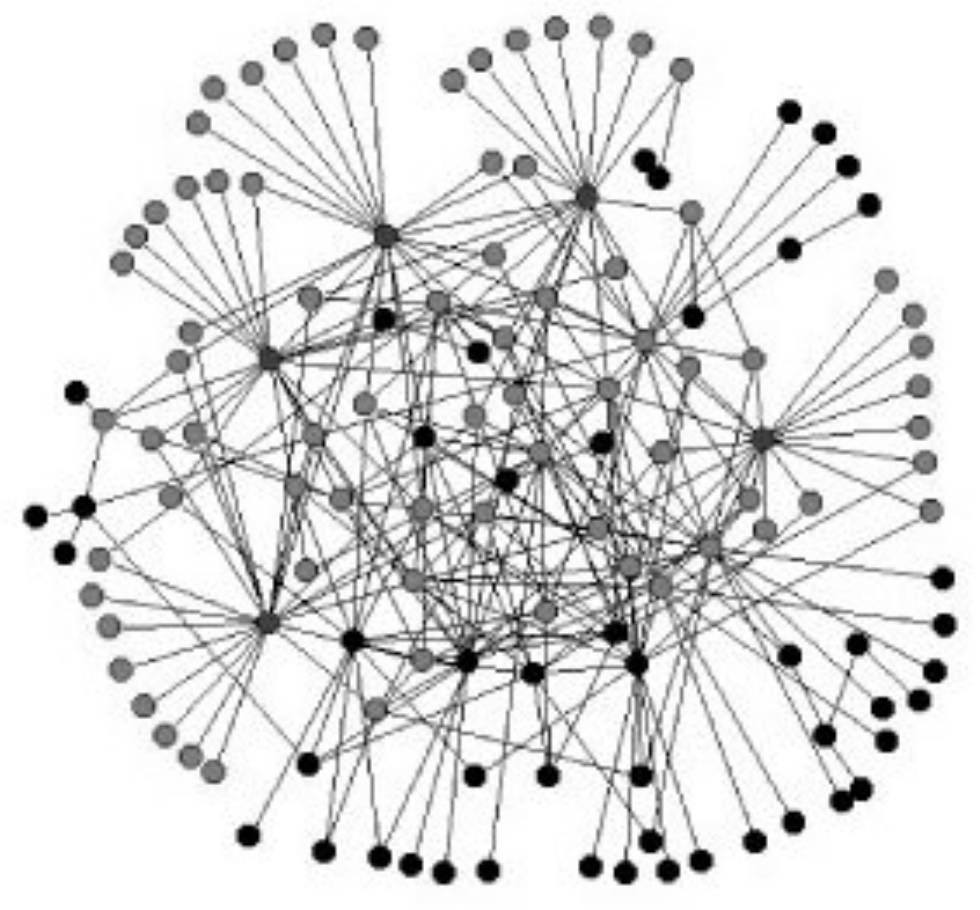}}
	\hspace{5mm}
\subfigure
	{\includegraphics[scale=.4]{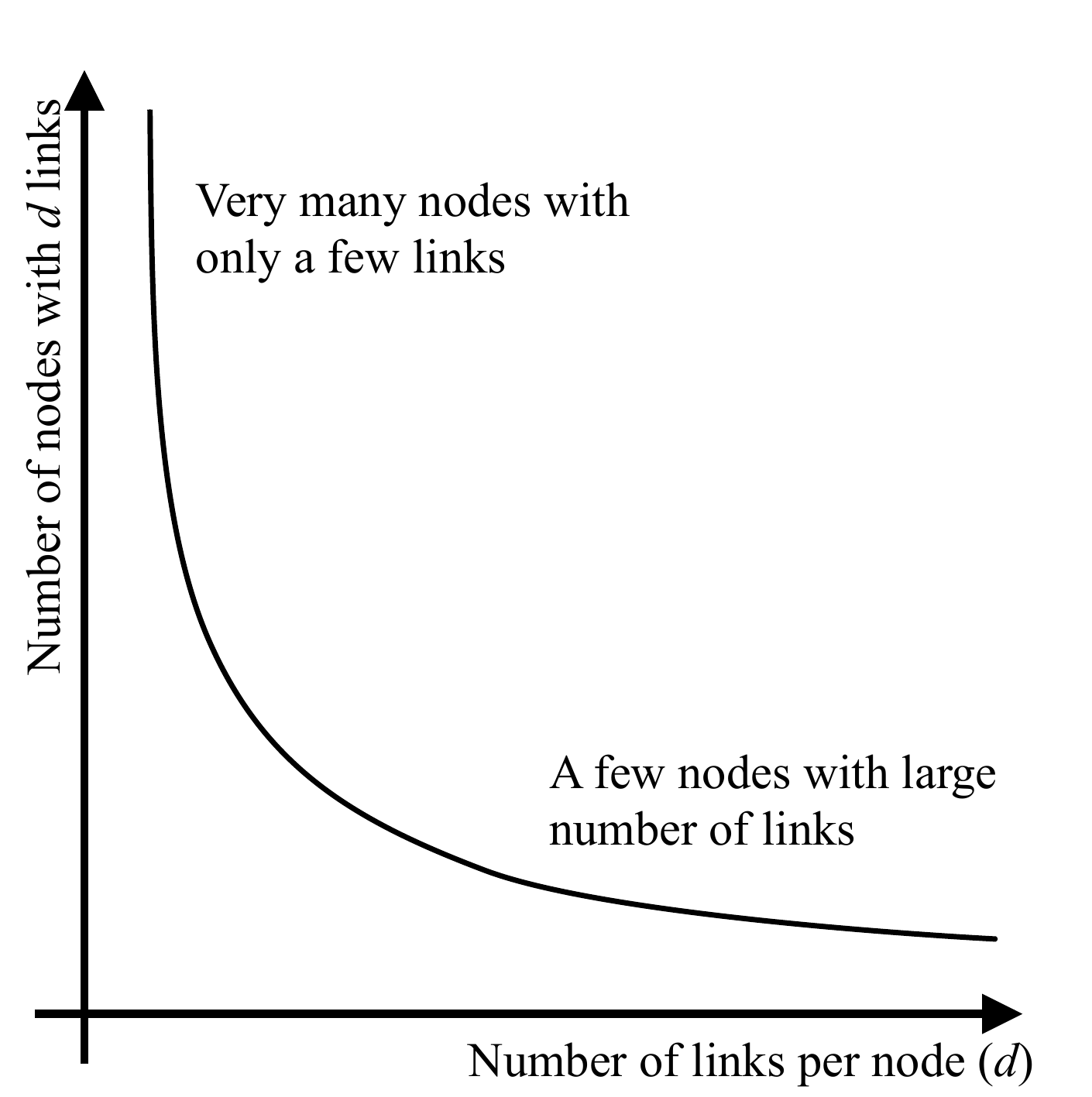}}
\caption{A scale-free network~\cite{AJB} and a power law degree distribution trend example.}
\label{examples}
\end{figure}

Figure~\ref{examples} shows an example of a scale-free network and an example of a power law degree distribution trend.

The BA model asserts that growth and preferential attachments determine the self-organization of a network with scale-free structure. Actually, real networks constantly grow by adding new nodes that at each time-step join the network and link to other nodes already present in the system. New nodes tie preferentially to those that are more highly connected, following a ``rich get richer'' phenomenon~\cite{BA}.

The probability $\Pi(k_i)$ that a new node will be connected to an already existing node $i$ depends on the degree $k_i$ for the property of the {\em preferential attachments}:

$$
\Pi(k_i) = \frac{k_i}{\sum_{j=1}^{N}k_j}
$$

The drawback of generic networks is that in case of a node disappearance, all links related to such node will be removed disrupting some paths among remaining nodes, eventually increasing the distance among nodes or fragmenting the network in a set of small portion or creating isolated nodes. On the other hand, scale-free networks present a high degree of clustering that determines their stronger robustness. Robustness is also due to the inhomogeneous connectivity distribution since the majority of nodes has only few links and the probability to select nodes with small connectivity will be much higher. Removal of such small nodes does not alter the network path structure and has no impact on the overall network topology. Therefore, scale-free networks are {\em resilient} against random errors~\cite{Bar} since they decay shrinking but not falling because nodes have the characteristic of communicate even in case of high failure rates~\cite{AJB}. When the power law distribution has a degree exponent less than $\gamma = 3$, scale-free networks have the property to remain connected indefinitely. Over the value of $\gamma = 1$ a critical point exists in which networks start to fall apart and they are destroyed through removing all nodes.

However, in case of a targeted attack aiming at eliminating, from the beginning, the most connected nodes, such networks will be rapidly destroyed. This vulnerability to attacks is due to the high degree of connectivity that is maintained by a few highly linked nodes and represents the {\em Achilles' heel} of scale-free network~\cite{AJB}.

\subsubsection{Network synchronization}
\label{ns}

Scale-free networks exhibit several interesting dynamical phenomena; in particular, we focus on the synchronization motion of their dynamical elements~\cite{WC1}. The aim of the following part of this work is to present a qualitative and paradigmatic correspondence between the emergence of synchronization in a network and the establishment of the choreography phenomenon.

In complex networks, nodes are characterized by a collection of variables that identifies a {\em state}. Such variables tend to synchronize so over time, they asymptotically assume identical values. Moreover, due to the self-organization process of scale-free networks, their syncronizability holds even when new nodes are constantly added. 

Let us introduce $x_i = (x_{i1}, x_{i2},\dots, x_{in}) \in S$ as the state variables of node $i$ in a network of $N$ nodes where $S \subset \mathbb{R}^n$, $c > 0$ a given constant representing the linking strength and $\Gamma \in \mathbb{R}^{n \times n}$  a square matrix linking coupled variables. Moreover, we observe a matrix $A$ in which $a_{ij} = a_{ji} = 1$ if there is a connection between two generic nodes $i$ and $j$ $(i \neq j)$ and $a_{ij} = a_{ji} = 0$ otherwise. The state equation of the network is:

$$
\dot{\mathbf{x}}_i = f(x_i) + c \sum_{\substack{
j=1\\
j \neq i}}^{N}
a_{ij} \Gamma (x_j - x_i), \quad i = 1, 2, \dots, N
$$
\noindent
where $f(x_i )$ is the dynamic of a single node and is a locally Lipschitz function which map the domain of $S$ into $\mathbb{R}^n$.

The dynamic of a scale-free network is said to be asymptotically synchronized over time $t$ if:

$$
\mathbf{x_1}(t) = \mathbf{x_2}(t) = \ldots = \mathbf{x_N}(t) = \mathbf{s}(t), \quad \mbox{as }  t \rightarrow \infty
$$
\noindent
where $s(t) \in \mathbb{R}^n$ is a solution for an isolated node, i.e. $\dot{\mathbf{s}}(t)=f(s(t))$ and $s(t)$ is therefore an equilibrium point.

A complete dissertation on this topic is out of the scope of this paper and we refer to~\cite{WC2} and references therein for further reading.

The equilibrium point represents the {\em state of choreography} and the synchronization enables the homogeneity of state variables of the members involved in choreography.

Note that the $n$ state variables in $x_i$ can represent both elements characterizing the member itself (intra-) and elements of interaction between members (inter-). In this way, we can define $x_i$ as composed by $k$ intra-state variables and $l$ inter-state variables, with $n = k + l$. Moreover, $\Gamma$ still represents coupling relationships since intra-state variables and inter-state variables can influence each other.

\subsection{Network membership}
\label{nm}
Inter-organizational innovation networks require peculiar membership characteristics for the potential members that are attracted to join networks to get some benefits.
Network membership characteristics are expressed by means of {\em ontology} and {\em homophily} properties.

\subsubsection{Ontology}
\label{ont}

The term ontology is relevant in many fields as knowledge engineering and artificial intelligence. Several definitions and a wide range of applications are considered in different topics (e.g. see~\cite{CJB} and references therein). 

We consider the definition of ontology provided in~\cite{Stu}: ``{\em An ontology is a formal explicit specification of a shared conceptualization}''. Ontology is therefore:

\begin{itemize}
\item {\em Formal}: it communicates the intended meaning of defined terms, independent on social or computational context. Formalism is a complete set of definitions express as logical axioms and documented in natural languages~\cite{Gru}.
\item {\em Explicit}: it defines the design of decisions, concepts and constrains.
\item {\em Shared}: it regards a knowledge accepted by a group in which the members agree about the objects and the relations of such knowledge.
\item {\em Conceptualized}: it concerns an abstract, simplified view of the world that we wish to represent. Every knowledge base, knowledge-based system or knowledge-level agent is committed to some conceptualization~\cite{Gru}.
\end{itemize}

In this way, the state variable $x_i$ given for every member $i$, the matrix $\Gamma$ and the matrix $A$ above mentioned are parts of the conceptualization, are formal since they respect a mathematical modelization, are explicit because every member knows its composition and shared because each member commits to have a state variable identically composed following the coupling rules.

The conceptualization is completed with the domain of knowledge, characterized formally by the set of objects that can be represented and is defined as the {\em universe of discourse}. The set of objects and their relationships are reflected in a specific vocabulary representing knowledge~\cite{Gru}. In our context, ontology is described by defining the names of entities in the universe of discourse (i.e. members, relations, processes, etc.) with a semantic description of each term. Members recognize a certain rate of consensus about the knowledge domain of ontology and are committed with it if their actions are consistent with its set of definitions. Indeed, members act respecting the ontology and behave rationally to achieve the final outcome of the network.

\subsubsection{Homophily}
\label{hom}

Homophily is the tendency of nodes to link with others that are similar to them or, in other words, the members' attitude to associate and connect with similar~\cite{MSC}. Individuals in homophily relationships share common characteristics that make communication and relationships easier. Inter-organizational networks foresee similar interests among members that share innovation attitude, goals and believes.

In the literature, homophily in scale-free network is related to network topological aspects in terms of node similitude related to preferential attachment. It represents a relevant aspect leading the growth of networks. In particular, homophily refers to how the preferential attachment privileges the bond between new nodes and those having high number of nearest neighbours and high similitude jointly~\cite{AMVS}.

\begin{figure}[htbp]
\begin{center}
\includegraphics[scale=.9]{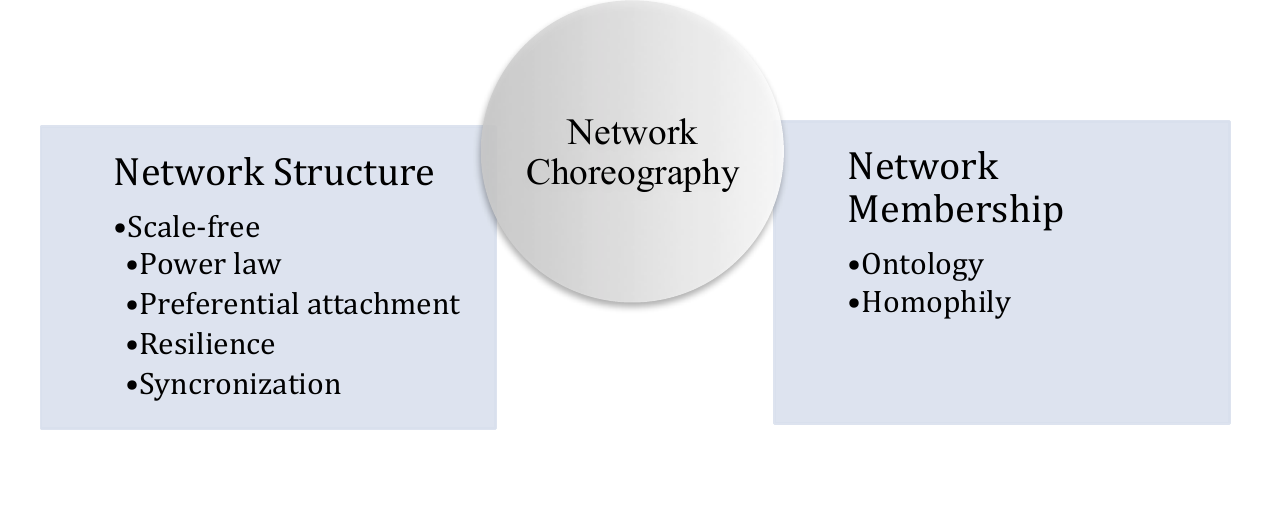}
\caption{Network properties.}
\label{np1}
\end{center}
\end{figure}

Figure~\ref{np1} resumes the network properties in our framework.

%%%%%%%%%%%%%%%%%%%%%% The case study %%%%%%%%%%%%%%%%%%%%%%
\section{The case study}
\label{EEN}

We introduce a case study to show the choreography characteristics of a real innovation network. Here, we deem the Enterprise Europe Network (EEN) that was launched in 2008 by the European Commission's Directorate-General for Enterprise and Industry. EEN's mission is helping small companies make the most of the business opportunities in the European Union (EU) by offering combined services according to the principle of one-stop shop for small businesses. Members of EEN are different and independent organizations as chambers of commerce, technology centres, universities, research institutes and development agencies.

In the case study, we study the collaborations among partners through the analysis of the Partnership Agreements (PAs) at country level. PAs comprise different services such as technological and commercial cooperation and collaborative research. 

In the following, a preliminary study of the EEN is presented to show the aptitude of the network to perform choreography. We remand to [13] for the complete analysis of the network and the in-depth study of the whole set of data integrated with a social network analysis.

The data collected by the Executive Agency of the EEN concern $52$ countries and $2019$ PAs signed during the period January 2011 - October 2012. The granularity of the available data is at country level, so the network is composed as follows: each node is a centroid that represents a country inside which there are independent organizations, as network partners, while links exist if two countries share at least one PA. 

In the paper, the analysis is restricted to the topology of the network since our aim is to sustain that the EEN posses the network properties described in the previous sections. In terms of network theory, we note a simple graph, i.e. a graph without multiple arcs between two nodes. Therefore, the number $M$ of arcs reflects interactions among countries, rather than the entire set of $2019$ PAs, thus decreasing the number to $M = 351$.

\begin{figure}[htbp]
\begin{center}
\includegraphics[scale=.8]{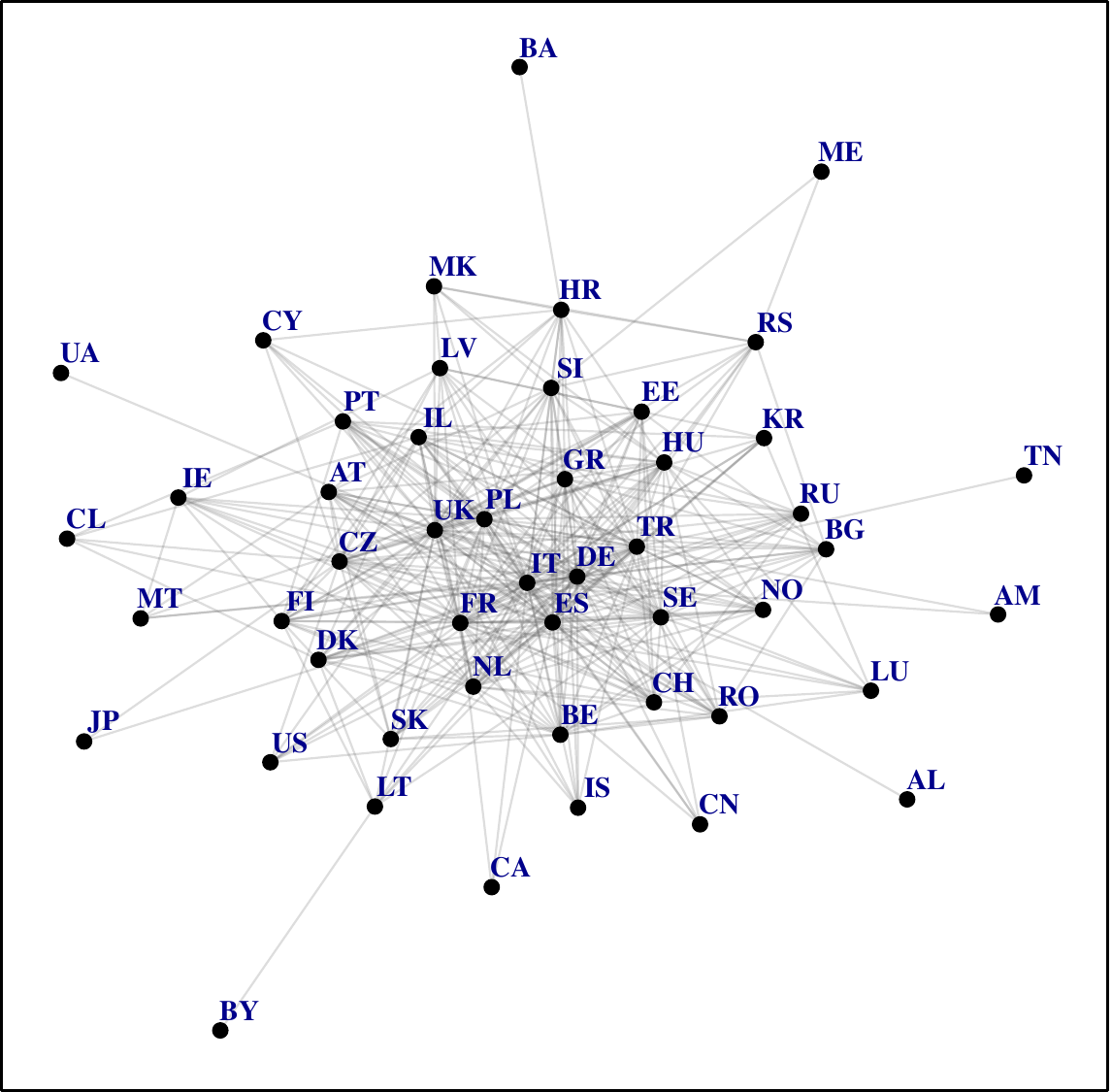}
\caption{The EEN network.}
\label{EENnet}
\end{center}
\end{figure}

The data processing and all the network analysis are conducted using the software R\footnote{http://www.r-project.org}  with the {\em igraph}\footnote{http://igraph.sourceforge.net} package. During the time period under observation, three EEN countries, namely Egypt, Mexico and Syria, have not signed any PA\footnote{Partners of EEN located in third countries operate under Article 21.5 of the Competitiveness and Innovation Program (CIP) and sign a Cooperation Agreement with the Executive Agency of the Network (EACI).}; hence, we do not consider them in the analysis. After this pre-processing, the network graph configuration is composed by $N = 49$ nodes and it is illustrated in Figure~\ref{EENnet} where node labels are the official country codes.

From the analysis of the network properties, we compute an average path length equal to $L = 1.82$ and an average clustering coefficient of $C = 0.66$.

\begin{figure}[htbp]
\begin{center}
 \begin{minipage}[t]{6cm}
   \centering
   \includegraphics[scale=.36]{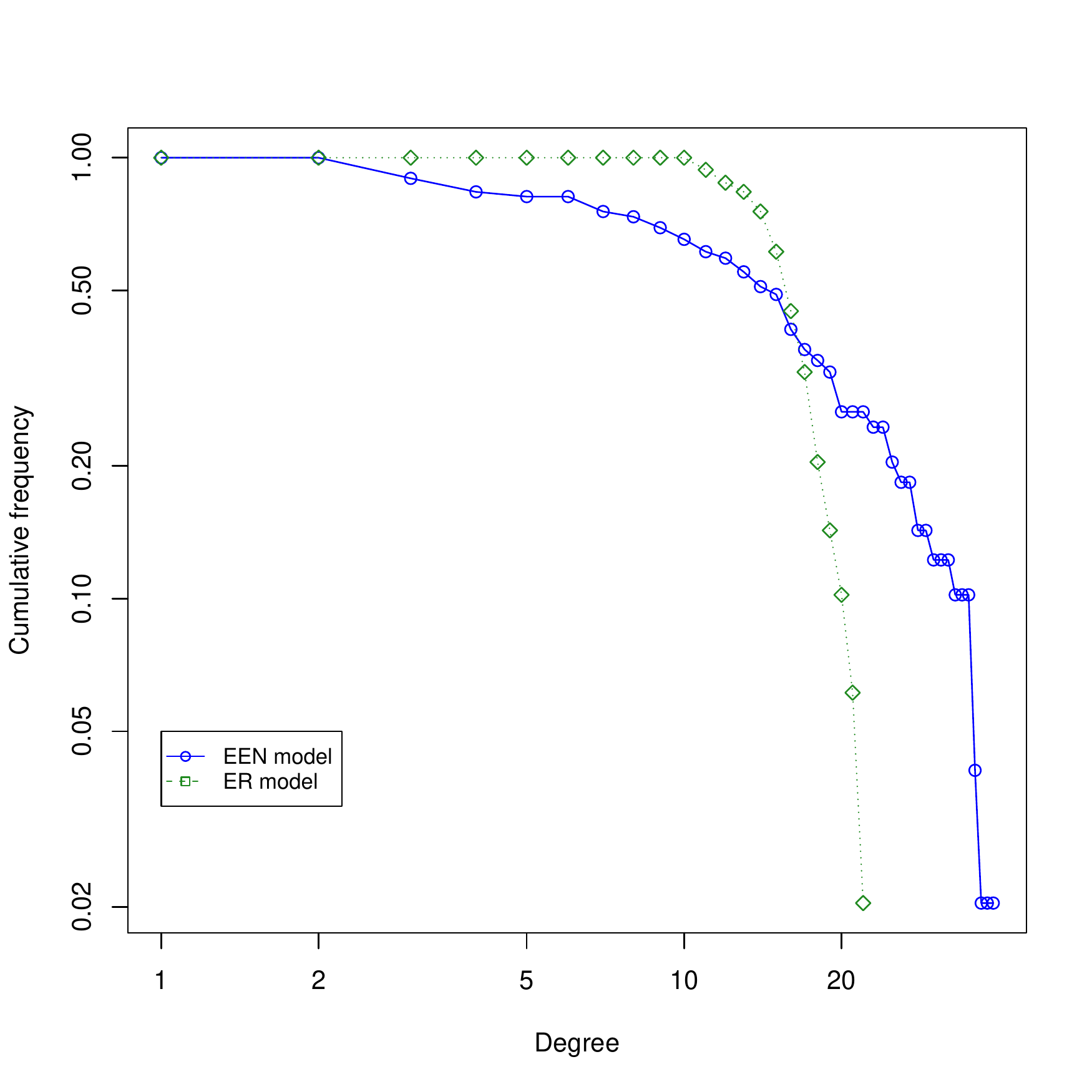}
   \caption{Degree distribution for the EEN network and the Erd\H{o}s and R\'enyi model.}
   \label{degreedist}
 \end{minipage}
 \ \hspace{2mm} \hspace{1mm} \
 \begin{minipage}[t]{6cm}
  \centering
   \includegraphics[scale=.33]{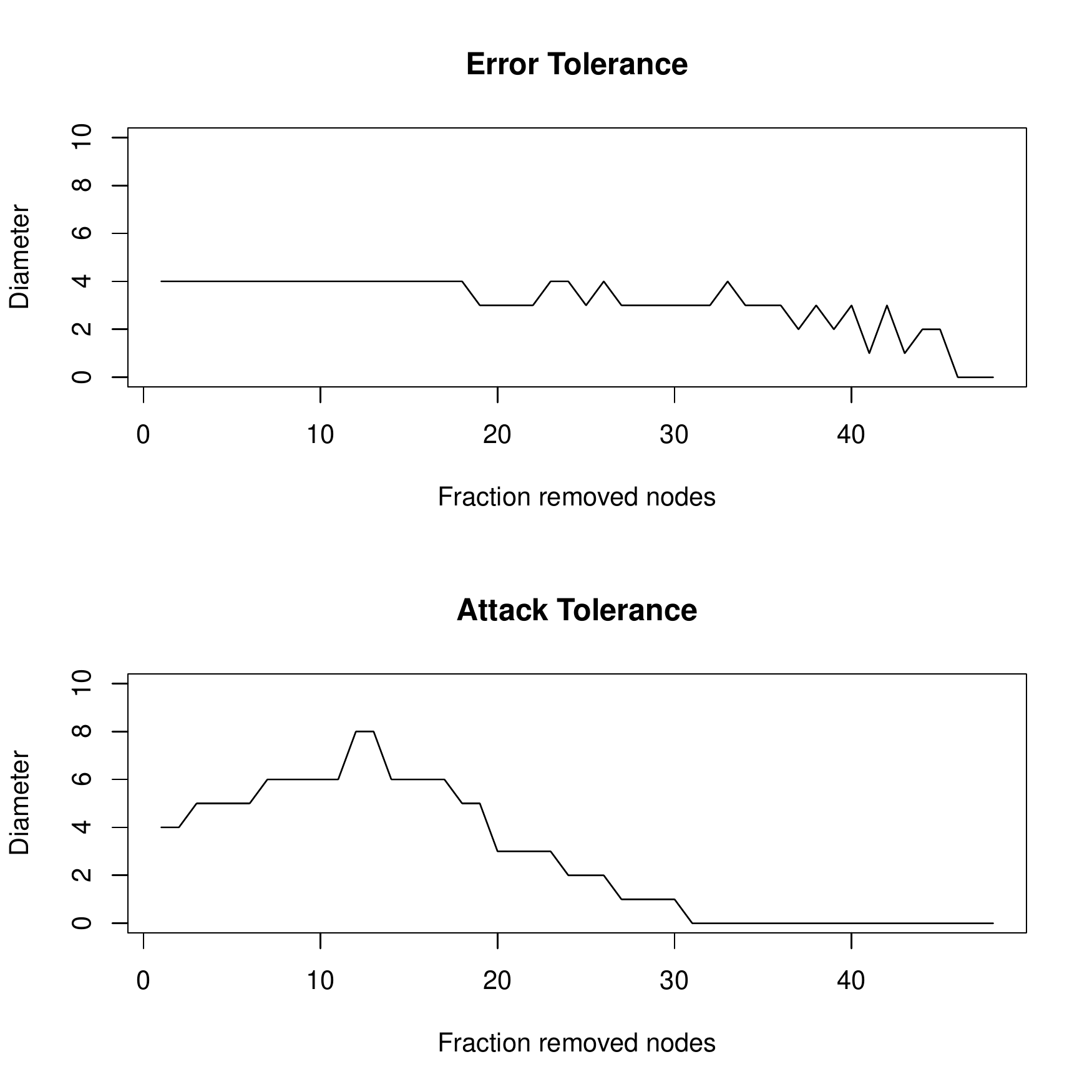}
   \caption{Network diameter under random error and attack.}
   \label{resilience}
 \end{minipage}
 \end{center}
\end{figure}

The power law degree distribution is tested by comparing the slope of the degree distribution of the network under observation and that of a random graph having the same number of nodes and links, and generated using the Erd\H{o}s and R\'enyi model. Figure~\ref{degreedist} shows the comparison between the two distributions. The power law slope of the EEN clearly decays less quickly than that of the random graph.

To evaluate the value of $\gamma$, we use the goodness-of-fit approach for fitting the power law distribution to data using a maximum likelihood estimator which results in a value of $\gamma  = 2.79$.

The values of the average path length, the average cluster coefficient and $\gamma$ meet the three scale-free properties required and they are comparable with the values of other real networks as reported in~\cite{WC2}.

As a consequence, we claim that the EEN exhibits the scale-free properties and is represented by the BA model.

As to the network resilience, we implement a double analysis, the Error Tolerance and the Attack Tolerance and their effects on the diameter of the network, i.e. the longest path between any two nodes. Indeed, a resilient network tends to retain its diameter at each time step a random node is removed, as in the case of error. The attack tolerance analysis is carried out removing at each time step the node with the highest degree, simulating an attack to the most connected nodes.

Figure~\ref{resilience} resumes the values of the diameter for growing values of the percentage of node removals and shows that the paths structure remains stable until about 80\% of node removals.

Table~\ref{tabella3} summarizes a collection of data for the EEN. In details, columns 1 and 2 report, respectively, the name of the country and its code, column 3 reports the degree $k_i$, column 4 indicates the clustering coefficient $C_i$ and columns 5, 6 and 7 show three basic measures for node centrality, namely closeness, betweenness and eigenvector centrality~\cite{WG}.

Closeness of a node gives higher score to a node having the short distance to every other nodes, betweenness gives higher score to a node that lays on many shortest path between other pairs of nodes and eigenvector centrality gives higher score to a node connected to many high score nodes.

The table highlights the presence of more than one hub since nodes as IT, UK, DE, ES, FR and TR have a neighbour value greater than 30; moreover, the majority of them has high centrality scores. The crosscheck between degree and centrality scores clearly shows the presence of the three classes of members, i.e. hubs, semi-peripherals and peripherals. There are some interesting cases as for HR and LT where a high score in betweenness corresponds to relatively small values in eigenvector centrality, meaning that these countries lay on many shortest paths but they are connected mostly to low score members.

\begin{table}[htbp]
%\begin{footnotesize}
\begin{tiny}
\begin{center}
\begin{tabular}{l|c|c|c|c|c|c}
\hline
               & \textbf{Country} &           &           &                     &                         & \textbf{Eigenvector} \\
\textbf{Country}  &  \textbf{code}     & \textbf{$k_i$} & \textbf{$C_i$} &  \textbf{Closeness}  &  \textbf{Betweenness}  & \textbf{centrality} \\
\hline\hline
Italy&IT&38&0.38&0.0172&168.31&1.00\\
United Kingdom&UK&35&0.45&0.0164&108.33&0.99\\
Germany&DE&34&0.47&0.0161&104.01&0.97\\
Spain&ES&34&0.48&0.0161&65.02&0.99\\
France&FR&34&0.43&0.0156&93.50&0.95\\
Turkey&TR&31&0.53&0.0152&48.76&0.94\\
Poland&PL&28&0.54&0.0147&34.87&0.86\\
Czech Republic&CZ&26&0.55&0.0139&45.41&0.80\\
Sweden&SE&26&0.60&0.0141&23.83&0.83\\
Hungary&HU&24&0.62&0.0137&19.27&0.78\\
Greece&GR&23&0.59&0.0135&22.39&0.73\\
The Netherlands&NL&23&0.67&0.0132&10.49&0.78\\
Belgium&BE&21&0.69&0.0130&9.35&0.72\\
Finland&FI&18&0.71&0.0125&7.57&0.62\\
Israel&IL&18&0.73&0.0125&10.35&0.64\\
Slovenia&SI&18&0.69&0.0127&37.10&0.62\\
Switzerland&CH&17&0.79&0.0122&3.86&0.63\\
Austria&AT&16&0.75&0.0123&6.63&0.58\\
Denmark&DK&15&0.77&0.0122&8.82&0.54\\
Croatia&HR&15&0.63&0.0120&51.73&0.48\\
Bulgaria&BG&14&0.85&0.0119&1.88&0.54\\
Estonia&EE&14&0.87&0.0118&1.21&0.54\\
Norway&NO&14&0.85&0.0118&1.48&0.54\\
Romania&RO&14&0.88&0.0118&0.97&0.54\\
Portugal&PT&13&0.73&0.0116&5.04&0.49\\
Russia&RU&12&0.94&0.0112&0.28&0.48\\
Slovakia&SK&12&0.89&0.0115&0.57&0.47\\
Latvia&LV&11&0.69&0.0114&2.34&0.38\\
Serbia&RS&11&0.73&0.0112&15.16&0.35\\
Ireland&IE&10&0.76&0.0110&2.72&0.36\\
Lithuania&LT&9&0.72&0.0111&47.15&0.33\\
South Korea&KR&9&0.67&0.0110&2.03&0.33\\
Iceland&IS&8&1.00&0.0110&0.00&0.35\\
Luxembourg&LU&8&0.75&0.0105&0.87&0.30\\
Macedonia&MK&7&0.90&0.0110&0.45&0.25\\
United States of America&US&7&1.00&0.0108&0.00&0.31\\
Cyprus&CY&6&0.87&0.0106&0.31&0.23\\
Chile&CL&5&0.60&0.0100&0.36&0.16\\
China&CN&5&1.00&0.0105&0.00&0.22\\
Malta&MT&5&0.60&0.0103&0.59&0.18\\
Canada&CA&3&1.00&0.0102&0.00&0.14\\
Armenia&AM&2&1.00&0.0097&0.00&0.09\\
Japan&JP&2&1.00&0.0093&0.00&0.08\\
Montenegro&ME&2&1.00&0.0081&0.00&0.05\\
Albania&AL&1&0.00&0.0095&0.00&0.05\\
Belarus&BY&1&0.00&0.0073&0.00&0.02\\
Bosnia and Herzegovina&BA&1&0.00&0.0077&0.00&0.02\\
Tunisia&TN&1&0.00&0.0092&0.00&0.05\\
Ukraine&UA&1&0.00&0.0093&0.00&0.05\\
\hline
\end{tabular}
\end{center}
\caption{Details of the EEN network.}\label{tabella3}
\end{tiny}
%\end{footnotesize}
\end{table}

The study of the synchronization can not be detailedly performed because not all state variables of each member are known. Nevertheless, in~\cite{WC1} it is sustained that if we contemplate the matrix $A$ of member adjacency modified with $a_{ii} = -k_i$, its largest eigenvalue is $\lambda_1 = 0$ and if the second eigenvalue $\lambda_2$ is negative and not close to $\lambda_1$, then the synchronization state is exponentially stable. 

In the case study, the EEN adjacency matrix has $\lambda_1 = 0$ and $\lambda_2 = -0.66$.  Therefore the synchronization is a stable state and, consequently, the choreography is a stable state for the EEN.

With respect to the ontology as a membership characteristic, we resume in Table~\ref{ontEEN} some basic features regarding the EEN. The set of rules, guidelines and common language are formalized, explicit, shared and conceptualized among network members. The features remain valid also for the mathematical model given by the adjacency matrix $A$, while it is worthwhile to mention that for the complexity of a real case it is virtually impossible to explicit the state variables and the matrix $\Gamma$.

\begin{table}[htbp]
\begin{center}
\begin{tabular}{ | l | p{9cm} |}
\hline
    \textbf{Ontology features} & \textbf{EEN} \\ 
    \hline\hline
    Formal & Network members are grouped in Consortia and are legally bound with the network executive agency by Framework Partnership Agreements.\\ 
    \hline
    Explicit & Operational manual and guidelines contain all the key information regarding the working practices. These documents contain the obligations and formalities that members should follow. \\ 
    \hline
    Shared & All rules and procedures are shared. A common language is accepted and used among members. The exchange of good practices is encouraged to spread knowledge, enhance excellence and professionalism across the network.\\
    \hline
    Conceptualized & Rules and guidelines are conceptualized. Members sign a ``code of conduct''.\\
    \hline
    \end{tabular}
    \caption{EEN ontology.}\label{ontEEN}
\end{center}
\end{table}

Concerning the homophily, in this case study nodes represent countries, therefore, this property is not particular meaningful as they are similar by definition. 

We remand to~\cite{FI2} for a detailed analysis of the network with more comments, while in this preliminary study, we prefer to focus on the compliance of the EEN to the network properties and membership required for the emergence of choreography.

%%%%%%%%%%%%%%%%%%%%%% Conclusions and final discussion %%%%%%%%%%%%%%%%%%%%%%

\section{Conclusions and final discussion}
\label{cfd}
This paper represents the first attempt to describe the emergence of choreography in inter-organizational innovation networks characterized by the presence of different kind of members. This is the main difference with respect to orchestration theories, which contemplate the presence of a single hub that manages all activities to control the final outcome. The case of a single hub emerging in the network is a special circumstance of choreography that collapses into an orchestrated system; for this reason, we refer to some aspects and similarities of such theories to describe our model.

We suggest the BA model as an adequate framework to represent real inter-organizational innovation networks as in this model the network structure and its evolution are strictly correlated. However, the BA model highlights some limitations that should be overcome to more realistically describe the emergence of choreography. Certainly, the assumptions that all nodes increase their degree at the same time and that the growth rate is determined by node degree are not always confirmed in real networks. Moreover, the nodes' ability to acquire connections does not only depend on preferential attachments and on the age of the node in the network but on different other abilities. A way to face this particular problem is to introduce the role of {\em fitness}~\cite{BB}, which is the capacity of nodes to acquire links at different rates. This property reveals the network's perception of the importance of a node respects to the others.

The network topology allows to affirming the existence of a synchronization state and its similarity to the phenomenon of choreography. Although this similarity is only paradigmatic, it permits to give some hints about the state stability as showed in the case study.

Further research on the resulting framework should be conducted to extend it in a broader range of contexts and networks. Nevertheless, some essential characteristics have been studied. In particular, network topology influences activities among members; scale-free networks are able to spread and uphold interactions among nodes; resilience concerns the network capacity to face changes in the economic environment; ontology provides a shared common vision of the network and homophily enables cohesive relationships among members.

The case study collaborates the proposed framework. The analysis of the data shows the presence of a stable state of synchronization that we define choreography, which ensures members to profit from the value-added of the network according to their connections: members join EEN to get benefits from the network itself and from the networking processes to improve their competitiveness and innovation capability.

\section*{Acknowledgements}

\noindent
Authors would like to thank the Executive Agency for Competiveness and Innovation, the Evaluation and Monitoring Unit for the support to data retrieving, Benedetto Scoppola for his encouragements and for many fruitful discussions as well as Eugenio Archontopoulos for his helpful comments.


\end{document}